\documentclass[letterpaper]{article} 
\usepackage{aaai2026}  
\usepackage{times}  
\usepackage{helvet}  
\usepackage{courier}  
\usepackage[hyphens]{url}  
\usepackage{graphicx} 
\usepackage{natbib}  
\usepackage{caption} 
\usepackage{algorithm}
\usepackage{algorithmic}
\usepackage{paralist}

\usepackage{framed}
\definecolor{shadecolor}{gray}{0.95}
\newenvironment{intquote}
  {\begin{snugshade*}\begin{quote}\small}
  {\end{quote}\end{snugshade*}}

\usepackage{booktabs}
\usepackage{cleveref}
\usepackage{newfloat}
\usepackage{listings}
\DeclareCaptionStyle{ruled}{labelfont=normalfont,labelsep=colon,strut=off} 
\floatstyle{ruled}
\newfloat{listing}{tb}{lst}{}
\floatname{listing}{Listing}
\title{Who Will Become the Next Senior?\\How Generative AI Erodes the Development Pathway in Software Engineering}
\author{
    Sumin Yu\textsuperscript{\rm 1},
    Taesup Moon\textsuperscript{\rm 1, \rm 2}\thanks{Corresponding author.}
}
\affiliations{
    \textsuperscript{\rm 1} Department of Electrical and Computer Engineering, Seoul National University\\
    \textsuperscript{\rm 2} ASRI / INMC / IPAI / AIIS, Seoul National University\\
    \{ysmsoomin, tsmoon\}@snu.ac.kr
}

\usepackage{bibentry}

\begin{document}

\maketitle

\begin{abstract}
Generative AI (GenAI) is reshaping software engineering, raising concerns about how the development pathway through which juniors become seniors is being eroded.
While macro statistics show a decline in junior hiring and controlled studies demonstrate the effects of AI on individual task performance, the \textit{mechanisms} through which GenAI reshapes early-career \textit{development} in real organizational and educational contexts have not been thoroughly examined.
Through 14 semi-structured interviews with juniors at the threshold of entering software engineering and senior software engineers in South Korea, analyzed using Reflexive Thematic Analysis, we reveal a foundational pattern of Absorption---GenAI redirects entry-level work into senior–AI workflows---and three consequences: (1) juniors losing the productive struggle through which expertise once developed; (2) the structural reproduction of this loss through collective normalization of GenAI use in university classrooms; and (3) the perceptual asymmetry between seniors and juniors that prevents either side from correcting these dynamics on their own. 
By extending learning theory and situated cognition to organizational and institutional scales, we argue that GenAI
appears to be absorbing
not just specific categories of tasks but also
parts of the pathway
through which the next generation of seniors is formed.
Preserving this pathway will require deliberate institutional design across classrooms, workplaces, and the evaluation criteria for juniors.
\end{abstract}

\section{Introduction}

The advent of generative AI (GenAI) has transformed software engineering at unprecedented speed.
Tools like GitHub Copilot, ChatGPT, and Claude now perform substantial portions of code generation, debugging, and documentation that were frequently assigned to entry-level engineers.
Concurrently, junior developer hiring has contracted sharply.
In the United States, entry-level postings at the fifteen largest U.S. technology firms fell 25\% between 2023 and 2024 \citep{brynjolfsson2025canaries}, and programmer employment declined 27.5\% between 2023 and 2025 \cite{rak2025ai}. In South Korea, major technology companies have largely suspended open recruitment of junior software engineers, shifting to experience-based ad hoc hiring \cite{chosunbiz2025itrecruit}.

At the individual level, GenAI clearly accelerates developer productivity \cite{brynjolfsson2025generative,peng2023impact,cui2026effects}.
At the collective level, however, this raises one underexamined question:
if entry-level work is no longer reaching junior engineers, who will become the next generation of seniors?
Nevertheless, these hiring trends are commonly read as job displacement---efficient substitution of human labor with automated tools (i.e., GenAI), accompanied by corresponding adjustments in headcount.
We do not dispute that this displacement is occurring.
We would argue, however, that the tasks currently being absorbed by GenAI, such as minor bug fixes, routine implementation, and debugging, are not simply \textit{entry-level tasks} that organizations can eliminate without further issues and concerns.
What GenAI is absorbing is the \textit{developmental scaffolding} through which juniors grow into seniors:
the early stages of an apprenticeship-based development pathway,
where juniors learn through real-world experience alongside more experienced developers, and expertise is passed down from one generation to the next \cite{lave1991situated}.
When this scaffolding is absorbed,
what may disappear
is not only entry-level employment but also the very \textit{mechanism} through which the next generation of seniors, \textit{i.e.}, skilled professionals, is formed.

We examine these dynamics through the case of South Korea, where this mechanism may operate strongly.
Korean workers have adopted GenAI faster than nearly any other national context---work-related AI usage reaches 51.8\%, almost double the U.S. rate \cite{bok2025korea}.
At the same time, open recruitment has collapsed, and Korea's labor market structures displace adjustment costs onto labor market entrants rather than incumbent workers \cite{bok2025youth}.
Therefore, rather than serving as a representative national case study, we note that South Korea functions as an important case in which these dynamics emerge earlier and more clearly.

Building on the macro-level hiring decline identified in numerous previous quantitative analyses, the contribution of our study lies not in confirming this phenomenon but in identifying its underlying mechanism and the factors contributing to its persistence.
We address three research questions through semi-structured interviews with 14 software engineers---eight juniors at the threshold of workforce entry, and six seniors with at least six years of industry experience:

\begin{compactitem}
\item (\textbf{RQ1}) What developmental capacities are juniors losing as GenAI restructures early-career work across workplaces and classrooms? 
\item (\textbf{RQ2}) Through what structural mechanism is juniors' productive struggle obstructed? 
\item (\textbf{RQ3}) Why do these dynamics persist without self-correction by individuals or organizations?
\end{compactitem}

Our contribution is threefold.
Empirically, we identify a foundational pattern of Absorption; GenAI redirects entry-level work into senior–AI workflows. From this pattern, three consequences follow.
First, juniors are losing the productive struggle through which expertise was once developed \cite{kapur2016examining,bjork1994memory}.
Second, this loss is reproduced as a structural condition through the collective normalization of GenAI use in university classrooms, foreclosing individual choice.
Third, the perceptual asymmetry between seniors and juniors \cite{lave1988cognition,brown1989situated} blocks coordinated self-correction.
Theoretically, we extend learning theory and situated cognition from the individual learner scale to the organizational and institutional scales where these dynamics actually occur.
Normatively, we argue that the conditions for the growth of juniors cannot be restored through individual effort alone; they must be re-established through deliberate institutional design.
\section{Background and Related Work}
\subsection{GenAI and the Changing Labor Market}

The rapid proliferation of GenAI tools, such as ChatGPT, Claude, and GitHub Copilot, has prompted growing concern about their impact on labor markets, particularly for early-career workers.
Global reports consistently show a structural shift in workforce demand,
including sharp declines in U.S. entry-level technology hiring \cite{brynjolfsson2025canaries}, as well as in overall programmer employment \cite{rak2025ai}.
These trends do not appear to be merely the result of economic cycles, and employers are increasingly redefining what they expect from new hires.
A large-scale analysis of approximately 11 million UK job postings from 2018 to 2024 found that demand for AI roles grew 21\% 
while university degree requirements for those same roles declined by 15\%, suggesting a structural shift toward skill-based hiring \cite{bone2025skills}.
Moreover, recent evidence further suggests that GenAI adoption is shifting work away from entry-level tasks across knowledge-intensive occupations, including software engineering, effectively narrowing the bottom stages of internal career ladders \cite{hosseini2025generative}.

South Korea offers a particularly clear lens on these dynamics.
Alongside its unusually high workplace GenAI adoption \cite{bok2025korea},
South Korea's open recruitment---in which major companies had long hired new software engineers in cohorts---has largely collapsed:
leading technology companies have shifted to experience-based rolling hiring and halted open junior recruitment \cite{chosunbiz2025itrecruit}, and IT-sector job postings decreased by 43\% between 2023 and 2024, with entry-level positions accounting for only 4.4\% of postings \cite{chosunbiz2025itrecruit}.
The Bank of Korea has framed this as
seniority-biased technological change, in which wage stickiness combined with patterns of seniority-biased adjustment concentrates costs on new entrants in the labor market rather than existing workers \cite{bok2025youth}.

However, it is important to note that establishing a causal link between GenAI adoption and the decline in junior hiring remains challenging.
Empirical analyses suggest that the employment conditions for AI-exposed entry-level jobs had already begun to weaken in early 2022---months before ChatGPT's public release---suggesting that broader economic shifts, post-pandemic corrections, and monetary tightening were key contributing factors \cite{brynjolfsson2025canaries}.
Nevertheless, the convergence of labor market shifts and the rapid adoption of GenAI is reshaping junior roles and the competencies they require, regardless of how much of the broader hiring decline can be attributed to GenAI.

While existing research has documented these macro-level shifts through labor market statistics and employer surveys, such analyses leave critical questions unanswered;
how do senior and junior engineers experience the changing role of early-career work, and where do their perceptions diverge?
Through what mechanisms is GenAI reshaping early-career development in software engineering, and why do these mechanisms persist, preventing juniors who have completed their education from entering the workforce?
This study addresses these gaps through qualitative interviews with \textit{both} groups.

\subsection{GenAI and the Changing Nature of Workplace and Education}
In addition to a broader impact on labor markets, GenAI is transforming the nature of work across many fields \cite{eloundou2024gpts}. In the realm of software engineering in particular, as software development practices change, there is growing interest in how developers can integrate GenAI into their day-to-day workflows \cite{liang2024large,barke2023grounded,vaithilingam2022expectation}.

In terms of productivity, studies consistently find that AI-powered coding assistants accelerate task completion for individual developers \cite{brynjolfsson2025generative,peng2023impact,cui2026effects}.
Longitudinal research on a software development organization \cite{xiao2025ai} further suggests that while AI tools have not resolved persistent team-level problems such as accountability for results or fragile communication, they have helped reshape team norms: by accelerating individual tasks such as coding and documentation, they establish efficiency as a baseline expectation and the responsible use of AI as a marker of professionalism.
Moreover, on the industry side, expectations of what new hires should bring are concurrently shifting \cite{wef2025futureofjobs}.
Where previous cohorts of junior engineers were expected to bring foundational coding skills and develop further on the job, employers increasingly expect new entrants to arrive with AI literacy as a baseline competency \cite{maslej2025aiindex}.
In South Korea, a Delphi survey of software industry experts \cite{shin2025impact} found that creative problem-solving ability, AI utilization capability, and domain understanding have emerged as the key competencies employers now seek, displacing the technical coding skills that were previously prioritized.
As a result, the nature of entry-level positions is being redefined, while the barriers to entry are becoming even higher.

On the other hand, a recurring concern is deskilling \cite{braverman1998labor}, i.e., the erosion of human expertise, which occurs when automated systems absorb tasks that previously required deliberate practice and skill development.
Recent research on GenAI revisits this concern at a new scale. People working in various knowledge industries anticipate that GenAI, even when framed as a tool for performing `menial work under human review,' may exacerbate the existing forces of deskilling, dehumanization, and disconnection \cite{woodruff2024knowledge}.
This concern is particularly acute for novice and early-career practitioners: a diary study of young professionals
found that ChatGPT use produced a diminished sense of ownership and challenge, captured in the question that titles the paper ---``If the machine is as good as me, then what use am I?'' \cite{kobiella2024if}.

Despite this growing body of work, most existing studies focus on individual productivity and team-level dynamics, and little empirical attention has been paid to how the introduction of GenAI is reshaping the \textit{development processes} through which junior engineers develop expertise in real organizational contexts---a gap this study seeks to address.

Additionally, there is research analyzing the adoption of GenAI from a student perspective.
A study of novice programmers, where the participants were students, found that while GenAI tools provided short-term productivity gains for some, they simultaneously widened the gap between those who struggled and those who progressed quickly by exacerbating pre-existing metacognitive difficulties and creating an illusion of competence---those who struggled often believed they had performed better than they had \cite{prather2024widening}.
Qualitative research on software engineering students documents recurring patterns of GenAI dependency and uncertainty about skill formation \cite{choudhuri2025insights}.
However, such studies have focused on student experiences in isolation, without incorporating the perspectives of industry seniors who can evaluate and hire these students.

Overall, existing research has largely examined this transition from either the educational side or the workplace side, not examining how those directly experiencing this structural change---such as junior engineers entering the workforce and senior engineers already within organizations---actually perceive it. This study contributes such an account.

The most recent research that is most similar to our research, which covers both juniors and seniors, is done by \citet{feng2026junior}: a recent mixed-methods study of agentic AI-mediated software engineering found that junior developers employed in industry oscillate between AI over-reliance and cautious avoidance and that agency in such work is shaped more by organizational policy than by individual preference.
However, a key distinction is that, in our study, the term `junior' refers to individuals preparing to enter the workforce, whereas in their definition, it refers to those who have already entered it.
By defining `junior' as ours, we can analyze the adoption of GenAI from \textit{both} the workplace and educational perspectives.

\subsection{Theoretical Foundations of the Development Pathway}

This developmental pathway has been theorized through the \textit{situated cognition} tradition.
\citet{lave1988cognition} argues that cognition cannot be isolated from the social settings and practices in which it takes place.
\citet{brown1989situated} extend this view:
practitioners' understanding of what counts as a problem or solution emerges from their direct engagement with practice, not from abstract reasoning that precedes action.
Closely related work by \citet{lave1991situated} develops the concept of \textit{Legitimate Peripheral Participation} (LPP): an apprenticeship-like trajectory in which newcomers join a community as recognized members and learn by gradually moving from peripheral, lower-stakes tasks to more central roles---competence develops through participation itself, rather than through instruction.
From this perspective, the developmental value of early-career work lies not only in the tasks themselves, but in the social and organizational context in which they are performed \cite{begel2008novice,latoza2006maintaining,dagenais2010moving,wenger1999communities}---learning occurs through doing, observation, and incremental responsibility.
The removal or reduction of lower-stakes entry-level tasks therefore represents not merely a change in workflow; it
disrupts the very structure through which expertise is reproduced.
This mechanism has been documented in other professional domains as well
\cite{beane2019shadow,beane2024skillcode}.

Research on the cognitive mechanisms of skill acquisition further explains \textit{why} removing peripheral tasks is consequential.
Kapur's theory of \textit{Productive Failure} posits that learning and performance are not commensurable: conditions that maximize short-term performance do not necessarily maximize long-term learning~\cite{kapur2016examining}.
Learners who first struggle with and fail to solve unfamiliar problems before receiving canonical instruction were found to have significantly greater conceptual understanding and application skills than those who receive instruction initially.
In other words, generating imperfect or failed solutions enables learners to recognize critical features of knowledge that structured instruction alone does not allow.
Bjork's related concept of \textit{desirable difficulties} similarly identifies conditions that impede short-term performance but enhance long-term retention and transfer~\cite{bjork1994memory}.
Crucially, both frameworks converge on a counterintuitive insight: smooth and successful performance during the early stages of learning is often a signal of superficial encoding rather than deep understanding.
Taken together, these frameworks suggest that the junior-to-senior developmental pathway depends on a structured, incremental exposure to authentic organizational tasks, beginning with lower-stakes, error-prone work and gradually increasing in complexity and responsibility. This pathway has served as the primary mechanism through which organizations reproduce expertise across generations.

The question raised by the present study is whether GenAI adoption---by absorbing the lower stages of this ladder---is disrupting this process in ways that existing organizational theory has not yet fully accounted for.
Our study examines this question through the lived experiences of both juniors preparing to enter the workforce and seniors currently working in the field, in South Korea, offering an empirical analysis of how these theoretical disruptions manifest in practice.

\section{Method}

\begin{figure}[t]
  \centering
  \includegraphics[width=\columnwidth]{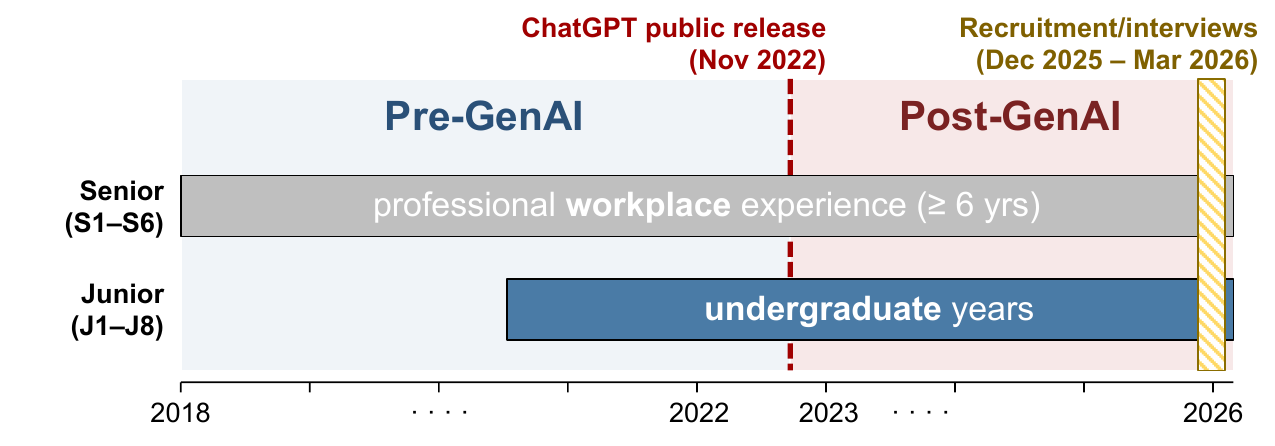}
  \caption{Temporal positioning of study participants.
  Juniors straddle the periods before and after the adoption of GenAI during their undergraduate years; seniors entered the industry before its adoption and continue to work in the field.
  }
  \label{fig:cohort-timeline}
\end{figure}

\subsection{Study Design}
We conducted a qualitative study using semi-structured interviews to investigate how the emergence of GenAI tools has reshaped skill expectations and career development pathways for early-career software engineers in South Korea. This study is detailed by three research questions (RQs), which are described in the Introduction section. Semi-structured interviews were chosen to enable participants to share their experiences while allowing the researchers to explore emerging themes in depth.

\subsection{Participants}
We recruited 14 participants through purposive sampling drawn from the authors' personal and professional networks. The participants were divided into two groups: eight juniors and six seniors.
Participants were selected based on their temporal positioning relative to GenAI adoption, illustrated in Figure~\ref{fig:cohort-timeline} and detailed below for each group.
In accordance with the principle of information power \cite{malterud2016sample}, we judged sample adequacy based on the specificity of our study population, the focus of our research objectives, and the analytic depth pursued.

\textit{Junior participants} (J1--J8) were defined as individuals within two years of completing an undergraduate degree, representing those at the threshold of entering the professional workforce. All junior participants were undergraduate students or recent graduates majoring in Electrical and Computer Engineering (ECE) or Computer Science (CS), with career trajectories oriented toward software engineering.
Notably, all junior participants experienced both the pre- and post-GenAI period during their undergraduate education, having taken courses both before and after the widespread availability of tools such as ChatGPT.

\textit{Senior participants} (S1--S6) were defined as professionals with at least six years of industry experience, drawn from diverse organizational contexts: two from large enterprises, two from mid-sized companies, and two from startups, spanning domains including game development, e-commerce, robotics, and others.
We purposefully designed this diversity to further observe how the adoption of GenAI varies depending on organizational size and technical domain.
Although senior participants were recruited from diverse organizational contexts, all held undergraduate degrees in ECE or CS, sharing the same educational foundation as the junior group, and are engaged in software engineering-related work in their respective fields.
Table~\ref{tab:participants} provides an overview of participant demographics.

\begin{table}[t]
\centering
\small
\setlength{\tabcolsep}{1mm}
\begin{tabular}{@{}lllll@{}}
\toprule
ID     & Years      & Organization Size  & Domain      & Major  \\
\midrule
J1--J8 & (student)  & N/A        & N/A         & ECE/CS \\
S1     & 9--11      & Large Enterprise & Electronics & ECE    \\
S2     & 9--11      & Large Enterprise & Game        & ECE    \\
S3     & 11--12     & Startup    & Robotics    & ECE    \\
S4     & 6--7       & Startup    & E-commerce  & ECE    \\
S5     & 9--11      & Mid-sized Company  & Game        & ECE    \\
S6     & 6--7       & Mid-sized Company  & Re-commerce & CS     \\
\bottomrule
\end{tabular}
\caption{Participant Overview. Junior participants (J1--J8) are current students or recent graduates; they have not yet entered the workforce, so Organization Size and Domain do not apply. Years indicates years of industry experience.}
\label{tab:participants}
\end{table}

\subsection{Data Collection}
Interviews were conducted in March 2026, in Korean, the native language of both the participants and the authors.
Each interview lasted approximately 50 minutes and audio recordings were made for all interviews with participants' consent.
Ethical approval was granted by the Institutional Review Board (IRB) of Seoul National University (IRB No. 2603/004-009).
All recordings were transcribed using an automated speech recognition tool (CLOVA Note\footnote{https://clovanote.naver.com/}), followed by manual correction by the authors.
Since the authors are native Korean speakers, data analysis was conducted using the original Korean transcripts; quotes appearing in this paper are English translations of selected Korean excerpts, reviewed by the authors for accuracy and idiomatic fidelity.
\subsection{Interview Structure}

Separate semi-structured interview guides were developed for junior and senior participants, each tailored to their respective roles and experiences.
The junior guide focused on perceived skill gaps between university education and industry expectations, use of GenAI tools in academic and work contexts, and expectations about career entry.
The senior guide focused on changes in task delegation and mentoring practices since the adoption of GenAI, perceptions of junior competencies, and hiring criteria changes. Both guides were designed to elicit participants' perspectives on the evolving relationship between GenAI capabilities and human skill development. Detailed interview question scripts for junior and senior are described in the supplementary materials.

Although the two guides differed in group-specific framing, they were designed around shared anchor topics---the changing nature of entry-level work, the conditions of skill formation, and expectations about junior–senior development---to which each group could respond from its respective vantage point.
This shared topical structure enabled cross-group analysis while maintaining the specificity of the questions asked in each group.
During analysis (described below), themes were constructed only when supported by accounts from both groups, except when asymmetry between groups itself constituted a meaningful analytic finding, as in our analysis of misaligned perspectives.

\subsection{Data Analysis}
All transcripts were anonymised prior to analysis by replacing participant names and organisational identifiers with coded labels (e.g., J1, S1, and Enterprise A).
We analyzed the data using Reflexive Thematic Analysis (RTA) \cite{braun2006using,braun2019reflecting}, an inductive approach that is well-suited to exploratory qualitative research seeking to understand participants' perspectives and experiences. We note here that in line with the theoretical orientation of RTA, we did not consider data saturation as a criterion for determining sample size \cite{braun2021saturate}.

Analysis proceeded through six phases:
(1) familiarization with the data through repeated reading of all transcripts;
(2) generating initial codes by annotating meaningful segments across the dataset;
(3) constructing candidate themes by grouping related codes;
(4) reviewing themes against the coded data and full dataset;
(5) defining and naming themes;
and (6) producing the final analysis.

Throughout our analysis, an audit trail was maintained to document how initial codes developed into final themes, supporting the transparency of the analytical process \cite{lincoln1985naturalistic}.
Disconfirming cases were
sought to ensure that the themes reflected a wide range of participants' perspectives, rather than only confirming instances.

\section{Findings}

\begin{figure}[t]
  \centering
  \includegraphics[width=\columnwidth]{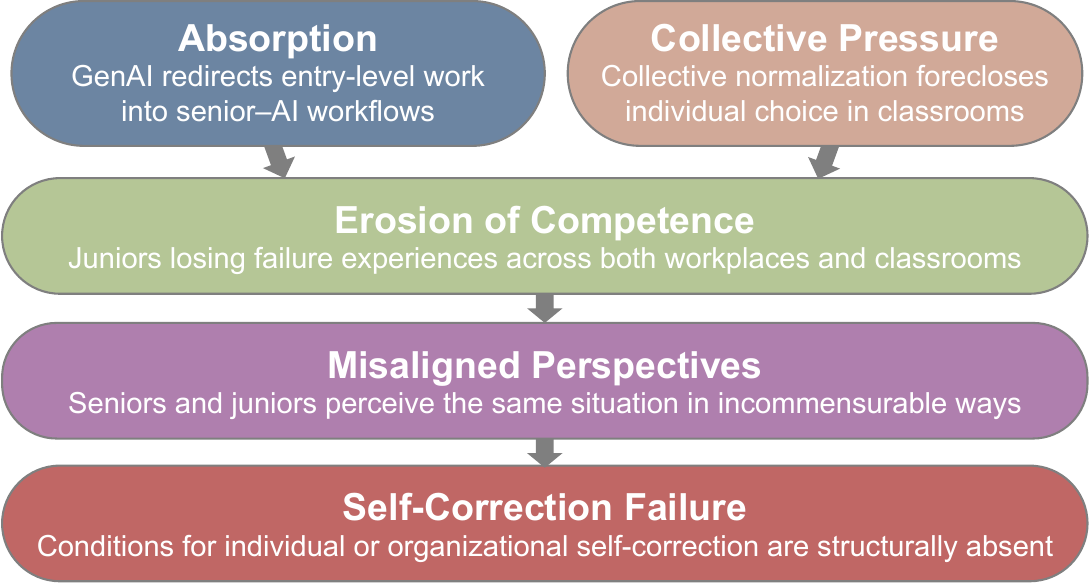}
  \caption{Overview of the paper's argument.
    Absorption and collective pressure (RQ2) produce the erosion of competence (RQ1); misaligned perspectives prevent self-correction (RQ3).
  }
  \label{fig:mechanism}
\end{figure}

Having introduced the methods of our study, we present our analysis under four themes.
The first---Absorption of Junior Opportunities into Senior Workflows---establishes the foundational pattern: GenAI expands senior capacity such that entry-level work no longer reaches juniors.
Three subsequent themes address our research questions---\textit{Erosion of Competence through Lost Failure Experiences} (RQ1) documents what juniors are losing across workplaces and classrooms; \textit{Collective Pressure as a Barrier to Productive Struggle} (RQ2) examines why this loss is structurally reproduced rather than individually avoidable; and \textit{Misaligned Perspectives Across Experience Levels} (RQ3) explains why these dynamics resist self-correction.
In each section, we present quotes alongside the corresponding analysis results to illustrate our claims; unless otherwise noted, the patterns these quotes represent were observed across multiple participants.

\subsection{Absorption of Junior Opportunities into Senior Workflows}
First, at the beginning of the interview, we examined how the emergence and adoption of GenAI has affected senior participants in their workplaces.    
A foundational pattern observed across senior participants was that GenAI tools had dramatically expanded their individual capacity, enabling them to handle tasks that would previously have been delegated to others, particularly entry-level engineers.
This had a direct structural consequence: the entry-level work that had historically provided \textit{learning opportunities} for entry-level engineers was no longer reaching them.

S4, a start-up founder with over twelve years of experience, described this shift in operational perspective.
Having significantly reduced his junior headcount over the last year, he observed that their departure had had no measurable impact on productivity.
As he put it,
\begin{intquote}
    ``We let a lot of junior engineers go in the second half of last year. But there was no impact whatsoever.''    
\end{intquote}\noindent
Following his remark that it was unavoidable
for
the company’s productivity, when asked what junior engineers could contribute that AI tools could not, S4 was blunt:
\begin{intquote}
    ``What can a junior engineer do better than a 100,000-won Claude subscription? I don't think there's anything.''    
\end{intquote}\noindent
This perspective---namely, junior employees were evaluated against the economic calculus regarding the possibility of being replaced by AI---was not unique to S4.
Among senior participants (S1--S6), it was consistently observed that AI tools had effectively absorbed more routine, lower-complexity tasks, e.g., code generation, initial debugging, and documentation. 
As S5 observed,
\begin{intquote}
    ``These days, things are being replaced so quickly that juniors are increasingly not being given the opportunity to do those things".    
\end{intquote}\noindent
On the other hand, one account made this pattern more complicated. S1, who worked at a large enterprise, explained how, as a senior, he had become able to handle most of the tasks previously done by juniors, but at the same time, he noted the following:
\begin{intquote}
    ``from the junior's perspective, the work they do and what they learn hasn’t really changed much. The only difference is that AI now exists.''
\end{intquote}\noindent
This was the only such observation across our senior interviews. In contrast to the absorption pattern described by other seniors, S1's account suggested that juniors in his organization continued to do similar work, with AI added as a supplement rather than as a replacement.
Rather than reading it as a simple outlier, we consider it as indicating boundary conditions for the absorption mechanism.
The seniors who described absorption most directly---notably S4 in a startup, where ``we let a lot of juniors go in the second half of last year''---operated in settings where task allocation is filtered primarily through efficiency calculus.
S1's large enterprise context represents the opposite pole, where institutional structures such as formalized onboarding pipelines and structured training programs determine the allocation of work, regardless of immediate productivity considerations.
While our sample cannot confirm this organizational difference as the underlying mechanism,
the contrast between S4's startup and S1's large enterprise suggests a tentative hypothesis:
absorption operates most freely where institutional structures for junior development are weak or implicit.
We return to this point through S6's onboarding design example in our Discussion section.

Based on the observations of the \textit{absorption}, we continued to interview seniors and, by analysing their responses alongside those of juniors, we identified the main themes that will be discussed in the next section.
 The following two themes examine what juniors lose when absorption proceeds without discernment, and how such losses manifest across the workplace and educational sites that engage in their skill development.

\subsection{Erosion of Competence through Lost Failure Experiences}

A central finding for both junior and senior participants was that, as they began using GenAI whether in school or at work, \textit{failure} no longer served as a source of learning.
Specifically, as GenAI absorbed the lower-complexity tasks that once characterized early-career workflows, the \textit{productive struggle} embedded in those tasks also disappeared.

All junior participants (J1--J8) noted that, compared to the pre-GenAI period, they can now achieve good results easily without the struggle they once experienced when completing university assignments or preparing for exams, and described what remained in its place as follows:
J6 said
``I don't know what not to do. I've only seen the good examples, so I don't know what the bad ones look like,''
and J1 described a more fundamental metacognitive issue, saying
``I don't know what I don't know.''
J2 elaborated on the same dynamics, saying:
\begin{intquote}
    ``If AI gives you the answer directly, the time you spend struggling gets shorter---so when you're faced with a problem, you might not even recall that a relevant concept exists.''    
\end{intquote}\noindent
As such, there was a consistent pattern among the participants regarding what emerged in place of \textit{failure: output without understanding}.
\citet{anthony2021knowledge} documented this phenomenon in adjacent professions: when novices use analytical tools without interpreting their outputs---what he terms `black boxing'---they cannot trace how outputs are generated, and therefore cannot learn from them.
J8 provided the most illustrative example, recalling a project presentation at her internship where she used unverified, AI-generated content:
\begin{intquote}
    ``There were things I was presenting that I didn't understand myself. I was thinking, 'Is this even right?'--- and I hadn't even checked, because there was no time.''    
\end{intquote}\noindent
Taken together, these accounts reflect what Kapur's theory of \textit{Productive Failure} would suggest: when learners are shielded from the imperfect, effortful attempts that they must endure to achieve mastery, they not only lose knowledge, but also the ability to recognize the boundaries of their own competence \cite{kapur2016examining}.
Similarly, Bjork's concept of \textit{desirable difficulties} suggests that the conditions that appear to slow progress, such as struggling, failing, and revising, produce durable and transferable learning \cite{bjork1994memory}.
GenAI, by eliminating these conditions, yields what J4 described as a distinct difference in his experience of two university courses covering the same subject.
\begin{intquote}
    ``My grades came out the same. But the growth I felt---I definitely studied much more in the machine learning course. The deep learning course just felt like I got better at using GPT.''    
\end{intquote}\noindent
The students, the academic field, and the evaluation criteria are all the same; the only difference is whether the assignment can be completed using GenAI.
The grade-equivalent outcomes are what make the case analytically significant: they show that the institutional signal of learning (the grade) can remain stable while what is actually learned diverges sharply.
In one course, the work produced durable domain knowledge; in the other, it produced fluency in just operating a tool, i.e., GenAI.
This case demonstrates the mechanism Kapur and Bjork would predict under nearly controlled conditions and
suggests that,
here, the current assessment structure in education
fails to identify it.

Even as growth through failure is fading as GenAI appears in the educational context, those failures remain an important element based on seniors' experiences and perspectives.
S6 explained what matters for professional development:
\begin{intquote}
    ``The reason someone is called a senior is that they know how work should \textit{not} be done, but now juniors can't learn from the bad examples.''    
\end{intquote}\noindent
Notably, junior and senior participants, standing at opposite ends of the experience spectrum, used nearly identical phrasing, ``knowing what should not be done'', to describe the missing competency.
Moreover, this perspective redefines \textit{seniority} not merely as the accumulation of correct solutions but also as a process of acknowledging failure through effort and challenging experience.
S3 further confirmed this from personal experience, describing his own decision-making skills nowadays as something built through years of incremental, hard work from the time he was in his early career:
\begin{intquote}
    ``Going through things one by one, making mistakes---the distress of having been wrong, accumulated over a very long period of time.''    
\end{intquote}\noindent
Moreover, what people can do in workplaces fundamentally is related to the skills that can be formulated by this failure process.
S5, a senior engineer at a mid-sized company, described this point as follows:
\begin{intquote}
    ``The cost of code generation is so cheap by using GenAI, but the cost of code verification is so expensive.''    
\end{intquote}\noindent
Such verification---knowing what is wrong, what does not work---is the human skill that failure-based learning produces. Yet juniors, by being shielded from failures, are no longer developing it. GenAI accelerates output production while eroding the very experiences through which juniors would learn to evaluate that output.

\subsection{Collective Pressure as a Barrier to Productive Struggle}

Whereas the previous section describes
the loss of failure experiences eroding junior developers' competence,
this section examines how those experiences are being eliminated in the first place.
What we observed is that this is not due to individual choice, but rather stems from a structural dynamic that operates within the educational environment itself.

A consistent pattern observed across junior participants was that the use of GenAI in university coursework is no longer a matter of personal choice.
Rather, it had become the norm for most students,
meaning that individual students found it difficult to resist using GenAI, regardless of their own intentions.
While this collective normalization reflects South Korea's broader context of rapid AI adoption \cite{bok2025korea}, the specific dynamics observed in the university classrooms of the junior participants are best traced through their own accounts.

The account provided by J4 was the most direct illustration of this dynamic: having initially chosen to work through assignments independently, he described the moment when this became impossible.
\begin{intquote}
    ``Everyone except me was using GPT, so everyone except me was getting nearly perfect scores. I no longer had the freedom to work through assignments on my own, making mistakes as I went---so in the end, I had no choice but to use it too.''
\end{intquote}\noindent
The pressure J4 describes is structural. In other words, the grading curve itself, as well as the fact that peers are using GenAI, has become the mechanism of coercion.
J5 expressed the same norm: ``Not using it is just stupid.'' J7 framed it as anxiety:  ``If I don't use it, I feel like I'm falling behind on my own.''
Accounts from all junior participants (J1--J8), including these three accounts, represent different but convergent responses to the same underlying collective pressure, leaving little room for individual resistance.

Crucially, this pressure operated not only among students but also at the institutional level in their universities.
J3 noted that faculty had effectively overlooked this practice:
\begin{intquote}
    ``The professors know students are using ChatGPT anyway... I don't think there was a single professor who told us not to use it.''     
\end{intquote}\noindent
J6 explained the institutional logic behind this silence:
\begin{intquote}
    ``The courses that AI can already do well are the ones where the original learning objectives aren't being met anymore. They should ban it---but you can't stop everyone from using it.''    
\end{intquote}\noindent
Consequently, a structural gap has emerged between the decisions about using GenAI in learning and what students actually are doing, leaving neither students nor institutions able to maintain the conditions necessary for \textit{productive struggle}, which was introduced in our previous finding, thereby hindering the growth of juniors.

What makes this observation consequential for the skill development of juniors is precisely identified by the accounts of J5:
\begin{intquote}
    ``The process has to be somewhat inefficient---I think that's just what learning requires. But in a society that values achievement, falling behind by not using it seems certain.''     
\end{intquote}\noindent
Albeit without using the theoretical term, this describes the \textit{desirable difficulty} proposed by \citet{bjork1994memory}.
J5 recognized this---and used AI anyway, because the educational structure left no viable alternative.

Not all junior participants felt this pressure to the same extent.
J2 described his ongoing personal resistance, stating, ``I keep trying not to take those shortcuts,''
while J6 identified open-ended, self-directed projects as spaces where independent thinking was still possible.
These accounts suggest that structural pressure varied across learning situations.
However, J2 also acknowledged the limits of such resistance:
\begin{intquote}
    ``I'm trying not to, but I'm so pressed for time that I end up doing it anyway.''
\end{intquote}\noindent
Individual intention, in other words, was undermined by structural constraint, suggesting that collective pressure exerts a stronger influence than personal preference.

\subsection{Misaligned Perspectives Across Experience Levels}

The preceding themes documented how this developmental pathway is being structurally obstructed.
However, a critical question remains: why do these mechanisms persist without being resolved?
Senior participants were not indifferent to the difficulties of juniors, and many expressed concern.
We address this question by examining how junior and senior participants perceive the same phenomenon in different terms and how that difference becomes a barrier to resolution.

A consistent pattern across senior participants was that the current situation appeared manageable. However, underlying this perception was an implicit assumption.
Namely, they assumed that juniors who had accumulated direct, hands-on experience would develop into competent professionals, combining that experience with AI fluency, much as seniors themselves had.
S5 articulated this most explicitly:
\begin{intquote}
    ``The fact that a junior lacks domain knowledge---I actually think he can easily fill that gap using AI.''
\end{intquote}\noindent
Subsequently, in the hiring process, he indicated that the most important thing would be direct experiential learning:
\begin{intquote}
    ``Even if there are no results to show, I think I'd give a lot of credit to someone who has directly worked with real data.''    
\end{intquote}\noindent
What this framing does not account for is whether the conditions it assumes are actually in place where juniors are educated.
As documented in the previous two sections, the educational contexts in which juniors are currently developing are ones in which GenAI has become the default mode of completing assignments and coursework---precisely the contexts in which the direct, effortful experience that seniors value would \textit{otherwise} have been accumulated.

As noted earlier in our discussion of Absorption, S5 himself acknowledged that juniors no longer encounter the very experiences he treats as a prerequisite, observing that ``the problem is that we can't wait for that time anymore.''
Yet this recognition coexists with the earlier assumption that juniors with direct, hands-on experience will emerge. S5 does not appear to recognize that these two statements are in tension.
This means that, following the emergence of GenAI, the skills that seniors expect from juniors are extremely difficult to acquire given the pace and trends of society---including the educational environment.
Seniors themselves are aware of this fact yet continue to hold those expectations.

Juniors experienced this same reality from a position that made its consequences \textit{immediate}.
While seniors viewed the situation as difficult but manageable, 
juniors experienced it as a process in which the opportunity to make knowledge their own had been denied.
J2 described this directly:
\begin{intquote}
    ``The less time you invest, the less you get back---it feels like the core is hollowed out.''    
\end{intquote}\noindent
J4 articulated the same experience in terms of ownership:
\begin{intquote}
    ``It becomes a skill you can only demonstrate when GPT is there. In the end, that's not really mine---it's not something I can draw on when I'm completely on my own.''   
\end{intquote}\noindent
From the senior side, the same reality appeared in fundamentally different terms.
S4 articulated this:
\begin{intquote}
    ``We are in a position to evaluate AI, whether this output is right or wrong, because of twenty years of accumulated experience. The next generation won't be in that position. Therefore, we are fine.''    
\end{intquote}

This divergence is not simply a matter of differing opinions or experience levels.
Rather, it reflects a related strand of the situated cognition tradition introduced earlier \cite{lave1988cognition,brown1989situated}---the idea that one's position within a practice shapes what is perceptible and what remains invisible.
Specifically, seniors who have already crossed the developmental threshold perceive the current situation as one that they can handle with ease and that is straightforward to navigate.
In contrast, juniors, who are on the verge of crossing the threshold, encounter the same situation as one in which crossing it has become more difficult.
Each group sees what their position makes visible; what the other experiences remains, to a significant degree, outside their field of perception.

This asymmetry extends beyond the present moment. While several senior participants expressed confidence that the current disruption would eventually resolve, their optimism was not directed at the juniors currently navigating the transition, but at a future generation who will grow up with AI from the beginning. 
S2 drew this distinction explicitly:
\begin{intquote}
    ``The current juniors are a bit in-between, honestly.''
\end{intquote}\noindent
Yet in the same breath, S2 expressed confidence:
\begin{intquote}
    ``Kids who've been building things with AI since they were young will already be at a senior level.''    
\end{intquote}\noindent
Junior participants recognized that they were not the intended beneficiaries of this optimism.
J5 observed:
\begin{intquote}
    ``We've experienced both worlds. But the ones who are doing well these days have been using it since middle and high school---I don't think they'd even notice there's a problem.''    
\end{intquote}

The consequence of these asymmetries is that attempts to address this developmental disruption tend not to converge.
Those in a position to act---seniors and organizations---do not fully perceive the structural conditions that juniors are navigating. Those who experience those conditions most directly---juniors---are not in a position to change them. S4, who recognized the problem with unusual clarity, captured this impasse:
\begin{intquote}
    ``\textit{Who is going to become the next senior?} This is a serious problem, and nobody has a solution.''    
\end{intquote}\noindent
When the perception of the problem and the experience of the problem do not coincide, the conditions for self-correction are structurally absent.
\section{Discussion}

\subsection{Absorption and Unprotected Pathway}
According to our findings, the mechanisms through which GenAI affects the skill development pathways of juniors can be summarized as follows: not only has GenAI taken over the entry-level tasks that previously supported the growth of juniors in the workplace, based on senior participants' accounts, but the skills that seniors---who currently use GenAI in their work---believe juniors should possess are also being hindered due to juniors' use of GenAI in university, based on junior participants' accounts, thereby widening the gap between seniors and juniors.
We go one step further to argue that these mechanisms did not arise solely due to the emergence of GenAI, but rather that they have been added onto the \textit{already unprotected} development pathway of juniors.

Senior participants in this study described their own developmental trajectories not as the product of formalized training, but as the incremental accumulation of hands-on work that, in retrospect, was \textit{not} systematized \cite{begel2008novice}.
S1, reflecting on his own early-career trajectory, found that the source of his growth was the combination of willingly doing \textit{menial} tasks and undertaking \textit{self-directed} efforts to expand beyond those tasks:
\begin{intquote}
    ``[They saw] that I really did the minor work diligently, exactly as told, without a word of complaint... And beyond that---since I didn't know anything about the technical work at the time---I would look up tutorials on my own and try to expand what I was capable of. When people saw that, they began entrusting me with more challenging tasks, and [my senior] came to trust me and gave me harder things, so my scope kept expanding. Looking back, I think it wasn't a matter of 'this is just minor work'---it was the mindset that I could do even that kind of work, and that's what allowed me to keep growing.''    
\end{intquote}\noindent
S4 echoed this pattern from his own trajectory in software engineering, recalling that the same accumulation of menial entry-level tasks had constituted his own developmental path. Working as a part-time developer \textit{without} specific assignments, he simply took on whatever came up:
\begin{intquote}
    ``When I was an undergrad, I started as a part-time developer at [Large Enterprise A], and the first thing I did was what we call the menial work of development. Writing code as told, debugging every day, checking and testing whatever came back... I did a lot of so-called development grunt work.
    [...]
    that's how I think I got to where I am now.''
\end{intquote}\noindent
Together, these accounts illustrate the unprotected character of this pathway in concrete terms: the conditions for development depended on individual circumstance, not institutional design.

The same S4, however, also reported that the contemporary equivalent of this work---namely, the entry-level tasks that juniors had previously been assigned in his startup, which had formed the very foundation of his cohort---had been absorbed by GenAI tools without any measurable productivity loss.
Yet, S4 describes this very work---the kind that had constituted his own developmental trajectory---as substitutable by a ``100,000-won Claude subscription.''
The contradiction doesn't seem to be a contradiction in his explanation: both statements appear obvious and make sense within their own contexts.
The absorption proceeds through general organizational rationales—efficiency calculations, headcount planning, productivity assessments.
As S4’s remarks suggest, no one needs to realize that this process is effectively blocking the pathways through which they themselves grew.
This lack of awareness is made structurally possible because the costs of adjustments arise at a different point in time and for a different group than the decision-makers.
In this way, the absorption continues, with no one in the right position to recognise or protect the pathway; what was already unprotected is further eroded.

\subsection{Extending Learning Theory and Situated Cognition}

\textit{Productive failure} \cite{kapur2016examining} and \textit{desirable difficulties} \cite{bjork1994memory}---identifying struggle, error, and revision as the conditions of deep, transferable learning---were articulated at the scale of the individual learner; our findings suggest that the conditions these theories identify as essential for learning are determined by a much broader dimension---such as the organizational and institutional levels.
Specifically,
our findings extend these theories to two larger contexts---the workplace and university classroom---where, respectively, the elimination of error-prone tasks and the collective normalization of GenAI use \cite{choudhuri2025insights}
each
foreclose productive struggle.
As our findings showed, individual disposition is overridden by collective pressure under such conditions, as J2's account of being unable to resist time pressure made explicit.
Therefore, we suggest that productive failure be designed at organizational and institutional scales, through structural arrangements rather than through learner choice. 

However, just pointing out that these conditions are getting weaker due to the use of GenAI doesn't explain why this phenomenon continues to be left unaddressed, and this is where the idea of \textit{situated cognition} is helpful.
As applied in our analysis of misaligned perspectives, \textit{situated cognition} \cite{brown1989situated,lave1988cognition,lave1991situated} holds that what is perceptible to a practitioner is shaped by their situated engagement with practice.
Building on this, our findings illuminate a more specific dynamic: when the practice itself is undergoing rapid transformation, the perception of that transformation diverges asymmetrically across experience levels.
In other words, seniors who have already crossed the developmental threshold perceive the current moment based on their accumulated competence,
whereas juniors mid-crossing the threshold perceive the same moment as one in which the threshold itself has receded.
Each group identifies what they see from their own perspective, as the contrast between S4's and J2's accounts in our findings is made explicit.
As a result, individual-level interventions, such as mentorship, advice, and encouragement, cannot, on their own, overcome this asymmetry. 
Combined with the organizational scale of the erosion itself described earlier, this implies again that sustaining the conditions for junior development requires structural intervention rather than individual remediation.

\subsection{Implications for Institutional Design}

If self-correction does not occur on its own, the conditions for junior development must be re-established through institutional intervention rather than relying on individual initiative.
S6's account illustrates what such intervention can look like in practice.
She described how her team integrates newcomers into active work, through what they call ``first step, one step, big step.'' 
The first step is small and concrete: ``finding something like a missing period in one of our production projects, putting the period in, committing it, opening a pull request, getting reviewed, and pushing it all the way to production''\footnote{All of these are simple and clear tasks for software engineers.}---that is, taking even the smallest change through the team's full product release process, so that the newcomer sees their own contribution become part of the final product.
Subsequent step expands into larger tasks juniors must approach with increasing self-direction---defining the problem themselves, consulting relevant seniors, and sharing what they have done---so that they can naturally learn how work progresses within the team as well as the technical workflow. The team assigns these tasks without explaining how to solve them, and in a final stage, they simply ask newcomers to ``find a problem and try to solve it.''
S6 frames this design as \textit{``a kind of learning by doing''}:
\begin{intquote}
    ``rather than teaching them `here's the process, learn it,' we've designed it so that they feel it by doing it themselves.''    
\end{intquote}\noindent
This approach concretizes one form of institutional intervention.
The key lesson from this case---namely, that the environment for training juniors should not be left to develop naturally but must be systematically designed---has significant implications across three areas: university education, evaluation criteria for juniors, and workplace onboarding.
At the educational stage, the question is how to preserve courses in which the learning objectives cannot be fulfilled by AI on the student's behalf---an issue J6 raised.
By revising educational standards to designate such courses as mandatory, ``reducing AI dependency'' can be established as a criterion for assessing educational quality rather than leaving a matter to faculty's discretion.
At workforce entry, we suggest that hiring criteria and evaluation could shift: as seniors are now recognized for their ability to solve higher-order problems using GenAI, evaluations for juniors might focus more on their capacity to recognize their knowledge gaps and errors when using GenAI.
Once juniors are inside the workplace, S6's example suggests that organizations may consider preserving designated learning spaces.
In other words, these spaces are not merely a way to improve work efficiency, but rather a place where employees can perform specific, small-scale tasks that are valued as a stepping stone for their growth.

What forms such interventions might look like can be motivated by examining two other fields that show similar patterns:
in aviation, FAA SAFO 13002 urges airlines to preserve opportunities for manual flying so that pilot skills do not atrophy as cockpit automation expands \cite{faasafo13002},
a concern confirmed by studies of cognitive degradation in manual flight \cite{casner2014retention,ebbatson2010relationship}.
In nuclear power, NRC requalification requirements mandate that licensed operators undergo periodic simulator training to maintain emergency-response capabilities that day-to-day automated operation rarely exercises \cite{nrc10cfr55_59}.
In both cases, institutional design preserves developmental conditions that automation would otherwise erode.

Without a comparable design, the development pathway through which the next generation of seniors might be formed will continue to attenuate.

\section{Limitations}

Our findings should be read alongside several limitations of the study's design and scope.
Our sample of 14 participants---eight juniors and six seniors---is appropriate for the kind of exploratory mechanism analysis we conducted (Braun and Clarke 2006, 2019), but does not support claims about the prevalence or magnitude of the patterns we identified. Nevertheless, we aimed to identify and interpret the mechanisms through which GenAI is reshaping junior development pathways, not to estimate their distribution.

Our cross-sectional design captures senior and junior perspectives at a single moment; the senior–junior comparison we present therefore reflects positional difference under conditions of situated cognition asymmetry rather than longitudinal change. Tracing how today's juniors actually develop over time will require future longitudinal work.

Our findings are situated in South Korea---a context characterized by the recent collapse of structured graduate recruitment, rigid employment protection, and one of the world's highest rates of GenAI adoption---and, as we note in the Discussion, whether the dynamics we documented emerge under different institutional configurations is a question for future comparative research.

Beyond sample size, our recruitment via purposive sampling through personal professional networks introduces selection effects. We deliberately constrained the sample to enable focused mechanism analysis: all participants were trained in electrical and computer engineering or computer science, and all junior participants had experienced both the pre- and post-GenAI periods during their undergraduate years. Details are demonstrated in our method section.
While this homogeneity supports the comparative analysis we conducted, there may be limitations to directly generalizing our findings to other professional fields, educational backgrounds, or cohorts that emerged either before or after the widespread adoption of GenAI.
\section{Conclusion}
Through 14 semi-structured interviews with juniors and seniors in software engineering, we have traced how GenAI is reshaping the early-career developmental pathway.
The mechanisms we identified---visible through participants' accounts
while operating beyond their individual choice---connect workplace task allocation and classroom norm formation.
The foundational pattern of Absorption
together with the structural mechanism of collective normalization in the university classroom
produces the erosion of competence through lost failure experiences across both workplaces and classrooms; and the perceptual asymmetry between seniors and juniors prevents self-correction through individual effort alone.
Crucially, this pathway was already unprotected; GenAI has made it consequential.
Deliberate institutional design in classrooms, in workplaces, and in the criteria recognizing junior competence will be required to form the next generation of senior practitioners.
The question in our title---"Who will become the next senior?"---names a responsibility that ordinary organizational rationality is now structurally incapable of discharging on its own.

\section{Acknowledgments}
This work was supported in part by the National Research Foundation of Korea (NRF) grant
[No. RS-2025-02263628], the Institute of Information \& communications Technology Planning
\& Evaluation (IITP) grants [RS-2021-II212068, RS-2022-II220113, RS-2022-II220959, RS-2021-II211343], and the BK21 FOUR Education and
Research Program for Future ICT Pioneers (Seoul
National University), funded by the Korean government. 

\section{Ethical Considerations}
This study was approved by the Institutional Review Board of Seoul National University (IRB No. 2603/004-009).
All participants provided informed consent prior to participation and were assured of confidentiality. Participation was voluntary, and participants were free to withdraw at any time without consequence. Audio recordings and transcripts are stored securely and will be retained for 3 years following publication.

\section{Positionality}
The authors are ML researchers with backgrounds in machine learning and AI systems, situated within the engineering tradition. This background provided sensitivity to the technical dimensions of GenAI capabilities and task boundaries, while requiring reflexive attention to potential biases toward technical explanations of sociotechnical phenomena \cite{berger2015now}. Throughout the analysis, the authors actively sought interpretations grounded in participants' own framings rather than imposing engineering-centric categories onto the data.

\section{Adverse Impact Statement}

This study documents structural dynamics through which GenAI is currently disrupting junior software engineer development. We anticipate three principal risks of misuse or misreading.
First, the empirical findings could be misread as confirming that junior engineers are economically substitutable by AI, providing cover for hiring decisions that further reduce entry-level opportunities. This reading inverts our argument: we document the absorption of developmental work as a problem to be addressed, not as a fact to be optimized. The mechanism we identify suggests that reducing junior hiring exacerbates rather than resolves the structural condition, by accelerating the disappearance of the development pathway through which the next generation of seniors is formed.
Second, our analysis of perceptual asymmetry could be cited to absolve senior engineers and decision-makers of responsibility---"seniors cannot perceive the problem because of situated cognition." The argument we make is the opposite: recognizing this asymmetry imposes a heightened, not diminished, duty of deliberate institutional design, because individual effort alone cannot restore the structural conditions that self-correction requires.
Third, our findings are situated in the South Korean institutional context, treated as a critical case in which several conditions converge. Direct policy transfer to contexts with different labor protection regimes, education systems, or AI adoption rates is not warranted. The mechanism we identify likely operates in those contexts but its strength, distribution, and points of remediation may differ; comparative empirical work is required before specific recommendations can travel.
We have framed our normative claims at the institutional rather than individual level precisely to discourage individualized interpretations---whether of blame (against juniors who use GenAI) or of optimism (that motivated individuals can resolve structural conditions on their own).

\bibliography{aaai2026}

\clearpage

\appendix
\section{Semi-Structured Interview Protocols}

The following guides reflect the semi-structured interview protocols used with junior and senior participants. They were developed around shared anchor topics---the changing nature of entry-level work, the conditions of skill formation, and expectations about junior–senior development---to which each group could respond from its respective vantage point. As semi-structured protocols, not all questions were asked verbatim or in fixed order; the order in which topics were taken up was adapted to each participant's responses and conversational style, and when accounts opened onto themes relevant to our research questions, follow-up questions were developed in the moment to deepen those threads.
Some questions used scenario-based framings---for example, asking senior participants to recall specific moments of considering AI versus junior delegation---to elicit concrete accounts rather than general opinions.
Representative follow-up probes are included below under some questions, but these were asked flexibly (e.g., with additional 
probes sometimes added) during the interview and did not follow a fixed schedule.
The original protocols were written and conducted in Korean; the translations below are provided for accessibility.

\subsection{Junior Interview Guide}
\subsubsection*{Opening / Background}
\begin{itemize}
    \item When did you first begin using GenAI tools in your studies?
    \item Could you describe the courses you have taken before and after the widespread availability of GenAI tools?
\end{itemize}
\subsubsection*{Reflections on your educational trajectory}
\begin{itemize}
    \item Looking back on your undergraduate experience, can you compare what learning was like before widespread GenAI availability versus after?
    \begin{itemize}
        \item \textit{Probe}: Were there specific courses, projects, or moments that crystallized this difference for you?
        \item \textit{Probe}: Have you noticed differences in your own study habits, sense of effort, or sense of accomplishment?
    \end{itemize}
    \item Have you experienced any course or project where the role of GenAI differed significantly, and what was that experience like?
    \item Are there capabilities you feel you developed primarily during the pre-GenAI period that you would not have developed in the same way today?
\end{itemize}
\subsubsection*{GenAI integration in academic work}
\begin{itemize}
    \item How often do you currently use GenAI in your learning process?
    \item Are there tasks where you feel that, because of GenAI, you have had fewer opportunities to actually experience the work yourself?
    \begin{itemize}
        \item \textit{Probe}: Could you describe a specific course or assignment where this was most pronounced?
    \end{itemize}
    \item Have you ever felt that you should be working through a problem on your own to develop your skills, but ended up relying on AI? How did you feel about that?
    \begin{itemize}
        \item \textit{Probe}: What pressures made it hard to resist using AI in that moment?
    \end{itemize}
    \item Has your way of studying changed over the past year, with LLMs becoming widely available?
    \item Looking at the topics you have learned so far---through coursework, projects, internships---are there parts where you think, ``GenAI could just do all of this''?
    \begin{itemize}
        \item \textit{Probe}: Does this realization change how you think about what you should focus on going forward?
    \end{itemize}
\end{itemize}
\subsubsection*{Anticipated transition to the workforce}
\begin{itemize}
    \item Among the topics you have studied in university, which do you think will directly apply to professional work?
    \item During your job search, have you noticed expressions or requirements in job descriptions that strike you as substantially different from what you have learned in school?
\end{itemize}
\subsubsection*{Hiring criteria and AI literacy}
\begin{itemize}
    \item Do you sense that what companies expect from new hires has changed compared to earlier periods? If so, how?
    \item In the era of GenAI, have you reconsidered what you should now be learning?
    \item How do you think you can demonstrate your AI competence to employers---through projects, portfolios, certifications, or other forms?
\end{itemize}
\subsubsection*{Closing}
\begin{itemize}
    \item Is there anything important about your experience with GenAI, your education, or your career development that we have not yet discussed?
    \item Is there anyone in your network---peers, recent graduates, mentors---whom you would suggest we speak with for this study?
\end{itemize}

\subsection{Senior Interview Guide}
\subsubsection*{Opening / Background}
\begin{itemize}
    \item Could you describe your current role, the organization you work in, and the team you work with?
    \item Have you been directly involved in training or mentoring juniors? Was this before or after GenAI tools became widely available---or both?
\end{itemize}
\subsubsection*{Reflections on your own developmental trajectory}
\begin{itemize}
    \item Could you walk us through your own development from a junior to a senior engineer? What kinds of work did you do during your first one to three years?
    \item Looking back, were there things you worked on as a junior that did not seem meaningful at the time but turned out to be important for your development?
    \begin{itemize}
        \item \textit{Probe}: Was there a particular moment of recognition or change for you during that process?
    \end{itemize}
    \item As a junior, were there tasks that a senior could have completed themselves but deliberately delegated to you for your development?
    \begin{itemize}
        \item \textit{Probe}: From the senior's perspective, what cost or risk did they take on by delegating these tasks?
    \end{itemize}
    \item In the team you currently work with as a senior, is junior development handled in a similar way to how it was when you were a junior---or has it changed?
    \begin{itemize}
        \item \textit{Probe}: If similar, why do you think these practices have persisted despite the broader changes in the field?
        \item \textit{Probe}: If different, when and why do you think this changed?
    \end{itemize}
\end{itemize}
\subsubsection*{GenAI integration and changes in team workflow}
\begin{itemize}
    \item How often do you currently use GenAI in your own work?
    \item Based on your own experience, what kinds of tasks does GenAI do well, and what kinds does it not do well?
    \item Have you recently had the experience of thinking, ``I could have delegated this task to a junior, but I just handled it myself with AI''? If you do not work closely with juniors, have you observed this pattern in other parts of your organization?
    \begin{itemize}
        \item \textit{Probe}: What was the deciding factor for not delegating (e.g., time, explanation, feedback cost, quality, or responsibility)?
    \end{itemize}
    \item Has the kind of work assigned to juniors changed since GenAI became available?
    \begin{itemize}
        \item \textit{Probe}: Have the criteria become higher? Has the nature of the work changed?
    \end{itemize}
\end{itemize}
\subsubsection*{Junior development pathway}
\begin{itemize}
    \item Earlier we discussed the kinds of work that have been most important for junior development. Looking at things now, who currently does that work? Has it disappeared from junior portfolios?
    \item Have you considered how these changes might affect the organization in the longer term?
    \begin{itemize}
        \item \textit{Probe}: Are you or your team thinking about how to address this?
    \end{itemize}
\end{itemize}
\subsubsection*{Hiring criteria and future competency needs}
\begin{itemize}
    \item In recent hiring of junior engineers, would you say the criteria are closer to ``they need to be more capable than before,'' or closer to ``we are now looking for different capabilities''?
    \begin{itemize}
        \item \textit{Probe}: What specifically are these new or different capabilities?
    \end{itemize}
    \item Do you think these capabilities can be adequately developed through university education?
    \begin{itemize}
        \item \textit{Probe}: If not, where do you think they should be developed?
    \end{itemize}
    \item Looking three to five years ahead, what competencies do you think will become most important for juniors?
\end{itemize}
\subsubsection*{Closing}
\begin{itemize}
    \item Is there anything important about GenAI's impact on junior engineering, team formation, or career development that we have not yet discussed?
    \item Is there anyone in your network whom you would suggest we speak with for this study?
\end{itemize}

\section{Analytical Approach}
We analyzed the dataset using Reflexive Thematic Analysis (RTA) following the framework articulated by \citet{braun2006using,braun2019reflecting}. RTA was chosen over alternative qualitative approaches---including codebook-style thematic analysis \cite{boyatzis1998transforming} and Grounded Theory---for two reasons. First, our research questions concerned the identification of mechanisms within an emergent phenomenon, not the construction of a new substantive theory from the ground up; this aligned with RTA's interpretive engagement orientation rather than Grounded Theory's theory-building program. Second, we are committed to the view that themes are constructed through analytical engagement rather than discovered through reliability procedures; this aligned with RTA's reflexive, single-analyst orientation rather than codebook approaches relying on inter-rater reliability.

Consistent with this orientation:
\begin{itemize}
    \item Analysis was conducted by the authors throughout, with iterative engagement across multiple readings of the full dataset.
    \item Inter-rater reliability scoring and saturation counts were not employed; \citet{braun2019reflecting} discuss why such procedures, drawn from different qualitative paradigms, are not constitutive of analytical rigor in RTA.
    \item Reflexive engagement was supported by the authors' positionality (described in the main paper), with explicit attention to potential engineering-centric biases in interpretation.
    \item Disconfirming cases (notably S1, examined as a boundary condition in Theme 1) were actively sought during phase 4 and integrated into the analysis rather than treated as outliers.
\end{itemize}

\section{Ethical Procedures}
\noindent\textbf{IRB approval.}
This study was approved by the Institutional Review Board of Seoul National University (IRB No. 2603/004-009).
The approval covered semi-structured interviews with software engineering professionals and students, audio recording, transcription, and analysis. No modifications to the approved protocol were made during the study.

\noindent\textbf{Recruitment.}
Participants were recruited through purposive sampling from the authors' professional network. Initial contact was made via direct messaging using a brief recruitment script that described the study's purpose, expected time commitment (approximately 50 minutes), voluntary nature, and anonymization procedures. 

\noindent\textbf{Informed consent.}
All participants provided informed consent prior to participation. Consent was obtained both verbally (recorded) and via written form in Korean, the participants' native language. The consent process explicitly covered: the study's purpose; what data would be collected and how it would be used; anonymization and storage procedures; the right to withdraw at any time without consequence; and the right to decline to answer specific questions. 

\noindent\textbf{Confidentiality and anonymization.}
Participant names and direct identifiers were replaced with coded labels (J1–J8 for juniors, S1–S6 for seniors) immediately upon transcription. Organizational identifiers were generalized to category labels (e.g., "Large Enterprise A," "Mid-sized Company"). Quotes selected for use in this paper were reviewed by the authors for any residual identifying details---specific project names, distinctive team sizes, technology stacks, or product references---and modified or omitted where re-identification risk was present.

\noindent\textbf{Data storage and retention.}
Audio recordings and transcripts are stored on a password-protected device accessible only to the authors. Identifying information (names, contact details, organization names) is stored separately from de-identified transcripts. All study materials will be retained for three years following publication, after which they will be permanently deleted in accordance with the IRB-approved retention plan.

\noindent\textbf{Handling sensitive topics.}
Several topics covered in interviews were professionally sensitive---for junior participants, reflections on faculty practices and anticipated job market entry under uncertainty; for senior participants, reflections on hiring decisions and team management changes. Interviewers explicitly reminded participants of the confidentiality protections at moments where sensitive material arose, and offered participants the option to mark specific statements as off-the-record. Statements marked off-the-record were not transcribed and are not represented in our analysis.

\noindent\textbf{Researcher positioning.}
The authors conducted interviews from a position of relative seniority with respect to junior participants and relative juniority with respect to senior participants. This dual positioning was attended to during interviewing---particular care was taken not to lead junior participants toward critical accounts of their education, and to avoid deferential acquiescence to senior accounts.

\end{document}